


\font\titlefont = cmr10 scaled\magstep 4
\font\sectionfont = cmr10
\font\littlefont = cmr5
\font\eightrm = cmr8

\def\ss{\scriptstyle}
\def\sss{\scriptscriptstyle}

\newcount\tcflag
\tcflag = 0  

\ifnum\tcflag = 0 \magnification = 1200 \fi  

\global\baselineskip = 1.2\baselineskip
\global\parskip = 4pt plus 0.3pt
\global\abovedisplayskip = 18pt plus3pt minus9pt
\global\belowdisplayskip = 18pt plus3pt minus9pt
\global\abovedisplayshortskip = 6pt plus3pt
\global\belowdisplayshortskip = 6pt plus3pt

\def\barsoff{\overfullrule=0pt}


\def\endignore{}
\def\ignore #1\endignore{}

\newcount\dflag
\dflag = 0


\def\monthname{\ifcase\month
\or January \or February \or March \or April \or May \or June%
\or July \or August \or September \or October \or November %
\or December
\fi}

\newcount\dummy
\newcount\minute  
\newcount\hour
\newcount\localtime
\newcount\localday
\localtime = \time
\localday = \day

\def\advanceclock#1#2{ 
\dummy = #1
\multiply\dummy by 60
\advance\dummy by #2
\advance\localtime by \dummy
\ifnum\localtime > 1440 
\advance\localtime by -1440
\advance\localday by 1
\fi}

\def\settime{{\dummy = \localtime%
\divide\dummy by 60%
\hour = \dummy
\minute = \localtime%
\multiply\dummy by 60%
\advance\minute by -\dummy
\ifnum\minute < 10
\xdef\spacer{0} 
\else \xdef\spacer{}
\fi %
\ifnum\hour < 12
\xdef\ampm{a.m.} 
\else
\xdef\ampm{p.m.} 
\advance\hour by -12 %
\fi %
\ifnum\hour = 0 \hour = 12 \fi
\xdef\timestring{\number\hour : \spacer \number\minute%
\thinspace \ampm}}}



\def\endtitle{}
\def\title#1\endtitle{\vskip.5in\titlefont
\global\baselineskip = 2\baselineskip
#1\vskip.4in
\baselineskip = 0.5\baselineskip\rm}

\def\endauthors{}
\def\authors#1\endauthors{#1}

\def\endabstract{}
\def\abstract#1\endabstract{\vskip .3in%
\centerline{\sectionfont\bf Abstract}%
\vskip .1in
\noindent#1}

\def\nopageonenumber{\footline={\ifnum\pageno<2\hfil\else
\hss\tenrm\folio\hss\fi}}  

\newcount\nsection
\newcount\nsubsection

\def\section#1{\global\advance\nsection by 1
\nsubsection=0
\bigskip\noindent\centerline{\sectionfont \bf \number\nsection.\ #1}
\bigskip\rm\nobreak}

\def\subsection#1{\global\advance\nsubsection by 1
\bigskip\noindent\sectionfont \sl \number\nsection.\number\nsubsection)\
#1\bigskip\rm\nobreak}

\def\topic#1{{\medskip\noindent $\bullet$ \it #1:}}
\def\endtopic{\medskip}

\def\appendix#1#2{\bigskip\noindent%
\centerline{\sectionfont \bf Appendix #1.\ #2}
\bigskip\rm\nobreak}


\newcount\nref
\global\nref = 1

\def\therefs{}


\def\ref#1#2{\xdef #1{[\number\nref]}
\ifnum\nref = 1\global\xdef\therefs{\item{[\number\nref]} #2\ }
\else
\global\xdef\oldrefs{\therefs}
\global\xdef\therefs{\oldrefs\vskip.1in\item{[\number\nref]} #2\ }%
\fi%
\global\advance\nref by 1
}

\def\listrefs{\vfill\eject\section{References}\therefs}


\newcount\nfoot
\global\nfoot = 1

\def\foot#1#2{\xdef #1{(\number\nfoot)}
\footnote{${}^{\number\nfoot}$}{\eightrm #2}
\global\advance\nfoot by 1
}


\newcount\nfig
\global\nfig = 1
\def\thefigs{} 

\def\figure#1#2{\xdef #1{(\number\nfig)}
\ifnum\nfig = 1\global\xdef\thefigs{\item{(\number\nfig)} #2\ }
\else
\global\xdef\oldfigs{\thefigs}
\global\xdef\thefigs{\oldfigs\vskip.1in\item{(\number\nfig)} #2\ }%
\fi%
\global\advance\nfig by 1 } 

\def\fig#1{\xdef #1{(\number\nfig)}
\global\advance\nfig by 1 } 


\newcount\cflag
\newcount\nequation
\global\nequation = 1
\def\eqlabel{(1)}

\def\nexteqno{\ifnum\cflag = 0
\global\advance\nequation by 1
\fi
\global\cflag = 0
\xdef\eqlabel{(\number\nequation)}}

\def\lasteqno{\global\advance\nequation by -1
\xdef\eqlabel{(\number\nequation)}}

\def\label#1{\xdef #1{(\number\nequation)}
\ifnum\dflag = 1
{\escapechar = -1
\xdef\draftname{\littlefont\string#1}}
\fi}

\def\clabel#1#2{\xdef\eqlabel{(\number\nequation #2)}
\global\cflag = 1
\xdef #1{\eqlabel}
\ifnum\dflag = 1
{\escapechar = -1
\xdef\draftname{\string#1}}
\fi}

\def\cclabel#1#2{\xdef\eqlabel{#2)}
\global\cflag = 1
\xdef #1{\eqlabel}
\ifnum\dflag = 1
{\escapechar = -1
\xdef\draftname{\string#1}}
\fi}


\def\eeq{}

\def\eqnn #1\eeq{$$ #1 $$}

\def\eq #1\eeq{
\ifnum\dflag = 0
{\xdef\draftname{\ }}
\fi 
$$ #1
\eqno{\eqlabel \rlap{\ \draftname}} $$
\nexteqno}







\def\eqa #1\eeq{
\ifnum\dflag = 0
{\xdef\draftname{\ }}
\fi 
$$ \eqalignno{ #1 } $$
\global\cflag = 0}


\def\ie{{\it i.e.\/}}
\def\eg{{\it e.g.\/}}


\def\cmp#1#2#3{{\it Comm.\ Math.\ Phys.} {\bf #1} (19#2) #3}

\def\npb#1#2#3{{\it Nucl.\ Phys.} {\bf B#1} (19#2) #3}
\def\plb#1#2#3{{\it Phys.\ Lett.} {\bf #1B} (19#2) #3}


\global\nulldelimiterspace = 0pt



\def\frac#1#2{{{#1} \over {#2}}\,}  
\def\hf{{1\over 2}}



\def\Dsl{\hbox{/\kern-.6700em\it D}} 
\def\dsl{\hbox{/\kern-.5300em$\partial$}}
\def\pxpsl{\hbox{/\kern-.5600em$p$}}
\def\ssl{\hbox{/\kern-.5300em$s$}}
\def\epssl{\hbox{/\kern-.5100em$\epsilon$}}
\def\delsl{\hbox{/\kern-.6300em$\nabla$}}
\def\lxpsl{\hbox{/\kern-.4300em$l$}}
\def\elxpsl{\hbox{/\kern-.4500em$\ell$}}
\def\kxpsl{\hbox{/\kern-.5100em$k$}}
\def\qxpsl{\hbox{/\kern-.5000em$q$}}
\def\sla#1{\raise.15ex\hbox{$/$}\kern-.57em #1}
\def\Pl{\gamma_{\sss L}}
\def\Pr{\gamma_{\sss R}}



\def\roughly#1{\mathrel{\raise.3ex\hbox{$#1$\kern-.75em
\lower1ex\hbox{$\sim$}}}}

\def\ol#1{\overline{#1}}





\def\Scd{{\cal D}}

\def\Scg{{\cal G}}

\def\Scl{{\cal L}}

\def\Scr{{\cal R}}


\def\ssl{{\sss L}}

\def\ssr{{\sss R}}

\def\ssv{{\sss V}}


\def\tr{\mathop{\rm tr}}

\def\det{\mathop{\rm det}}






\nopageonenumber
\baselineskip = 18pt
\barsoff


\def\veps{\varepsilon}

\def\psibar{\ol{\psi}}

\def\lft{\ssl}
\def\rht{\ssr}
\def\bk{\item{}}

\def\von{$O_\ssv(N)$}
\def\chon{$O_\lft(N) \times O_\rht(N)$}
\def\dps{\partial_+}
\def\Dps{D_+}
\def\sdps{\Scd_+}
\def\dms{\partial_-}
\def\Dms{D_-}
\def\sdms{\Scd_-}
\def\htDps{\widehat{D}_+}
\def\htDms{\widehat{D}_-}
\def\htsdps{\widehat{\Scd}_+}
\def\htsdms{\widehat{\Scd}_-}
\def\Scaa{a}
\def\Scll{L}
\def\Scrr{R}
\def\Scff{f}
\def\Scgg{G}

\def\WZW#1{\Gamma\left[#1\right]}


\rightline{March, 1994.}
\rightline{McGill-94/15, NEIP-94-001}
\rightline{hep-th/9403173}
\vskip .6in

\centerline{$\hbox{\titlefont Nonabelian Bosonization as
Duality}^\dagger$\footnote{}{${}^\dagger$
\eightrm Research supported by the Swiss National Foundation.}}
\bigskip

\vskip 0.25in
\authors
\centerline{C.P. Burgess${}^*$\footnote{}{${}^*$
{\eightrm Permanent Address:
Physics Department, McGill University,
3600 University St., Montr\'eal,
}}\footnote{}{\eightrm $\phantom{{}^*}$ Qu\'ebec,
 Canada, H3A 2T8. E-mail: cliff@physics.mcgill.ca.}
and F. Quevedo${}^{**}$\footnote{}{${}^{**}$ {\eightrm E-mail:
quevedo@iph.unine.ch.}}}
\vskip .25in
\centerline{\it Institut de Physique}
\vskip 0.05in
\centerline{\it Universit\'e de Neuch\^atel}
\vskip 0.05in
\centerline{\it CH-2000 Neuch\^atel, Switzerland.}
\endauthors

\vskip .3in
\vskip .25in

\abstract
Applying the techniques of nonabelian duality to a system
of Majorana fermions in 1+1 dimensions, invariant under a nonabelian
group \chon, we obtain the level-one Wess-Zumino-Witten model as
the dual theory. This makes
nonabelian bosonization a particular case of a nonabelian
duality transformation, generalizing our previous result for
the abelian case.
\endabstract


\vfill\eject

\section{Introduction}

\ref\witten{E. Witten, \cmp{92}{84}{455}.}
Since its first formulation by Witten \witten\ close to ten years ago
nonabelian bosonization has proven to be a very powerful tool for
analyzing two-dimensional fermionic field theories. Its central result
states the equivalence between a theory of $N$ majorana  fermions,
and a nonlinear sigma model whose fields take values in the group $O(N)$.

There is, however, a conceptual drawback to this equivalence as it
is usually
presented. The drawback is that the usual derivation is not
{\it constructive}.
The form of the required bosonic sigma model can be motivated from the
properties of the current algebra of the fermionic theory, but once such
arguments have been used to intuit its form, the main line of reasoning
is devoted to establishing the equivalence of the known fermionic and
bosonic theories. A more direct, constructive, procedure would start
{}from either theory and derive the other without any foreknowledge of its
form. Such a method would have the obvious advantage of lending itself
to potential generalization to other systems, for which the equivalent
theory is not already known.

\ref\basd{C.P. Burgess and F. Quevedo, preprint hep-th/9401105,
{\it Nucl. Phys.} B (to appear).}
\ref\duality{A. Giveon, M. Porrati and E. Rabinovici, preprint
RI-1-94, NYU-TH-94/01/01, hepth-9401139 (unpublished).}
\ref\doq{X. de la Ossa and F. Quevedo, \npb{403}{93}{377}.}
A first step in providing such a foundation for nonabelian bosonization
was
recently taken in Ref.~\basd, where it was shown how abelian bosonization
could be viewed as a special case of a wider class of techniques ---
collectively
known as {\it duality} transformations --- for proving relations among
quantum
field theories. This procedure  has become a well-defined
prescription for constructing an equivalent quantum field theory from
any given
one which has an abelian global symmetry
(for a recent review see \duality). The extension of this result to
nonabelian bosonization, using the recent extension \doq\ of the
dualization
prescription to theories with nonabelian symmetries, is the purpose
of the present
note. In so doing we intend to furnish a systematic and constructive
formulation
of the nonabelian bosonization technique.

\section{The Fermionic Theory}

Our starting point, as was the case for abelian bosonization, is the
fermionic
theory. We work in 1+1 spacetime dimensions, and take a theory of
 $N$ free
and massless  two-component majorana fermions, $\psi$. At the classical
level this theory enjoys  an \chon\ global flavour
symmetry under which the left-  and right- handed fermions rotate
amongst
themselves: $\psi \to (\Scl \Pl + \Scr \Pr) \psi$.  Only the diagonal
(vectorlike) \von\  subgroup, $\Scl=\Scr$, is anomaly
free, however, and so survives quantization.

For our present purposes, we imagine studying the correlations of the
 Noether
currents for the classical chiral \chon\ transformations. We may write
the
generating functional for these correlations in the following
way:\foot\conventions{Our conventions are: $\ss x^0 = t$,
$\ss x^1 = x$, $\ss x^\pm = {1\over\sqrt{2}}(x \pm t)$,
 $\ss \eta^{11}= -\eta^{00}= \veps^{01} =1$, $\ss \gamma_0 = i\sigma_1$,
$\ss \gamma_1 = \sigma_2$, $\ss \gamma_3 \equiv \gamma^0 \gamma^1
= \sigma_3$, and $\ss \gamma_{\lft} = \hf (1 + \gamma_3)$.}
\label\genfn
\eq \eqalign{
Z[\Scaa] &= \int [d\psi] \; \exp \left\{ -\,  {i\over 2} \int d^2x \;
 \psibar \,
\gamma^\mu (\partial_\mu - i \Scaa_\mu) \psi \right\} \cr
&= \int [d\psi] \; \exp \left\{ - \, {i\over 2} \int d^2x \; \psibar
\left[
\gamma^+ \Pr \Scd_+ + \gamma^- \Pl \Scd_- \right] \psi  \right\}. \cr}
\eeq
Here $\Scaa_\mu = \Scaa_\mu^a \, t_a$ are matrix-valued external fields,
with
$t_a$ being the generators of the \von\ symmetry. We take these generators
to be normalized according to $\tr(t_a t_b) = \hf \; \delta_{ab}$.
$\Scd_\pm \psi =  (\partial_\pm -i \Scaa_\pm) \psi$ similarly represent
the
background-covariant derivatives acting on $\psi$. Notice that the
 couplings
of the external fields promote the global classical chiral flavour
 invariance
to a local symmetry, provided
that they have the transformation rules: $\Scaa_+ \to \Scr \Scaa_+
 \Scr^\dagger
-i \partial_+ \Scr \, \Scr^\dagger$ and $\Scaa_- \to \Scl \Scaa_-
\Scl^\dagger
-i \partial_- \Scl \, \Scl^\dagger$.

\ref\pw{A.M. Polyakov and P.B. Wiegmann, \plb{141}{84}{223}.
}
\ref\wzwref{J. Wess and B. Zumino, \plb{37}{71}{95}.}
Such a system of free fermions is sufficiently simple to permit the
explicit
evaluation of the functional integrals \pw\ over the fermions. Since
 we
require this result later, we pause to record it here. Defining the
group-valued Wilson-line variables,
$\ell$ and $r$, according to $\Scaa_+ = i r^\dagger \dps r$ and $\Scaa_-
= i
\ell^\dagger \dms \ell$, as well as the field-independent constant $Z_0 =
Z[\Scaa =0]$, we have:
\label\answer
\eq
{Z[ \Scaa] \over Z_0} =  \left[ {\det \sdps  \; \det \sdms \over\det
\dps \; \det
\dms} \right]^{1/2} = \exp\left\{ -i \WZW{\ell r^\dagger} \right\}.
\eeq
The quantity $\Gamma$ --- {\it a.k.a.} the Wess-Zumino-Witten (WZW)
action \wzwref, \witten\ ---which appears in this equation
represents the following expression:
\label\WZWdef
\eq
\WZW{g} = {1 \over 16 \pi} \left[ \int_M d^2x \;
\tr\left( g^\dagger \partial_\mu g \; g^\dagger \partial^\mu g \right)
+ {2\over 3} \int_B d^3 x \; \veps^{\mu\nu\lambda}
\; \tr\left( g^\dagger \partial_\mu g \; g^\dagger \partial_\nu g \;
g^\dagger \partial_\lambda g \right) \right],
\eeq
where $M$ denotes the (1+1)-dimensional spacetime, and $B$ is a
three-dimensional region having $M$ as its boundary.

The symmetry of eq.~\answer\ under vectorlike background gauge
transformations is clear, given the transformation rules which $\ell$
and $r$ inherit from $\Scaa_\pm$: $\ell \to \ell \Scl^\dagger$ and $r \to
r \Scr^\dagger$. The properties of the WZW action are also such
as to ensure that eq.~\answer\ properly reproduces the fermion anomaly
for chiral \chon\ rotations.

%
Among the remarkable properties that are satisfied by $\Gamma$,
there is an identity, due to Polyakov and Wiegmann \pw, that
is particularly useful in what follows. This is:
\label\pwidentity
\eq
\WZW{gh^\dagger} =  \WZW{g} + \WZW{h^\dagger} - \, {1 \over
4\pi} \int_M d^2x \; \tr \left[ g^\dagger \dps g \; h^\dagger \dms h
\right].
\eeq

\section{Dualization}

\ref\rocekv{M. Ro\v cek and E. Verlinde, \npb{373}{92}{630}.}
In order to dualize this theory, we follow Refs.~\rocekv\ and \doq\
(see also
\basd) and promote the background \von\ invariance into a {\it bona fide}
gauge symmetry, by introducing a functional integration over its gauge
potential. We therefore
rewrite our path integral for $Z[\Scaa]$ in the following way:
\label\ggedgenfn
\eq \eqalign{
Z[\Scaa] &= \int [d\psi] \, [dA_\mu] \, [d\Lambda] \; \exp \left\{ i
\int d^2x \;
\left[ - \, \hf \; \psibar \, \gamma^\mu D_\mu \psi  + \veps^{\mu\nu}
\tr\left( \Lambda V_{\mu\nu} \right)  \right] \right\} \; \delta[\Scgg] \;
\Delta \cr
&= \int [d\psi] \, [dA_+] \, [dA_-] \, [d\Lambda] \; \exp \left\{ i
\int d^2x \;
\left[ - \, \hf \; \psibar \,  \left( \gamma^+ \Pr \Dps + \gamma^- \Pl
\Dms
\right) \psi \right. \right.  \cr
& \left. \left.
\phantom{\int [d\psi] \, [dA_+] \, [dA_-] \, [d\Lambda] \; \exp  i
\int d^2x \;
 - \, \hf \; \psibar }
+ 2 \,  \tr\left( \Lambda
V_{+-} \right) \right] \right\} \; \delta[\Scgg] \; \Delta, \cr}
\eeq
in which the covariant derivatives are defined with respect to both the
background field, $\Scaa_\mu$, and the new quantum field, $A_\mu$. \ie:
$D_\pm = \partial_\pm -i(\Scaa + A)_\pm$.

\ref\top{E. Alvarez, L. Alvarez-Gaum\'e, J. Barb\'on and Y. Lozano,
preprint CERN-TH-6991 (1993) (unpublished).}
Besides the quantum gauge potential, $A_\mu$, there are four other new
quantities in eq.~\ggedgenfn\ which need to be defined: $\Lambda$,
$V_{\mu\nu}$, $\delta(G)$ and $\Delta$.

\item{1.}
$\Lambda$ is a Lagrange multiplier field, which takes its values in the
Lie
algebra of \von. Its functional integration enforces the constraint that
$V_{\mu\nu}$ vanish. This constraint, together with the gauge condition
(more about which below) is chosen to have the unique solution
$A_+ = A_- = 0$,
in the absence of nontrivial spacetime topology.\foot\topissues{Some
topological issues arising in duality are addressed in Refs. \rocekv,
\top\ and \basd.} As a result, the integration over $\Lambda$ and
$A_\mu$ simply ensures that $A_\mu$ may be set to zero thoughout
the path-integral integrand, and this establishes
the equivalence of eq.~\ggedgenfn\ with the original fermionic theory,
eq.~\genfn.
\item{2.}
The tensor $V_{\mu\nu}$ is defined to be the difference
between the field strengths, $F_{\mu\nu}$ and $\Scff_{\mu\nu}$, that
are constructed from the two gauge potentials, $(\Scaa + A)_\mu$
and $\Scaa_\mu$. Explicitly: $V_{+-} \equiv F_{+-} - \Scff_{+-} =
\sdps A_- - \sdms A_+  -i \, [A_+, A_-]$. In using this as our constraint
we generalize slightly the procedure of Ref.~\doq, which uses the field
strength
built directly from $A_\mu$ alone. Our purpose in so doing is to keep
manifest the invariance with respect to vectorlike gauge transformations
of the background fields.  Notice that in order to have this invariance,
the Lagrange-multiplier field must acquire the transformation law
$\Lambda \to \Scg \Lambda \Scg^\dagger$, where $\Scg \equiv \Scl=\Scr$
is the
common group element for the background \von\ gauge transformations.
\item{3.}
Finally, the factor $\delta(\Scgg)$ is a functional delta function which
imposes an appropriate gauge condition, $\Scgg(x) = 0$, throughout
spacetime.
$\Delta$ represents the corresponding Fadeev-Popov-DeWitt determinant.
In what follows we will choose the background-covariant gauge, $\Scgg =
A_+ =0$, for which we may take $\Delta = 1$ (up to irrelevant
field-independent factors).

We now proceed to evaluate the functional integrals over $\psi$ and
 $A_\pm$,
leaving the Lagrange-multiplier field, $\Lambda$, as the bosonized
variable.
We do so in the following four steps.

\topic{1. The Fermion Integral}

The fermion integral may be directly performed using eq.~\answer.
In order
to use this expression, we require a definition of the Wilson-line
variables
for the  quantum field, $A_\mu$. We take: $(\Scaa + A)_+ =
 i (\Scrr r)^\dagger
\dps (\Scrr r)$ and $(\Scaa + A)_- = i (\Scll \, \ell)^\dagger \dms
(\Scll \, \ell)$.
Together with the previous definitions, $\Scaa_+ = i r^\dagger \dps r$
and
$\Scaa_- = i \ell^\dagger \dms \ell$, we therefore have $A_+ =
 i r^\dagger
(\Scrr^\dagger \dps \Scrr) r$ and $A_- = i \ell^\dagger (\Scll^\dagger
\dps
\Scll) \, \ell$. Clearly the new variables, $\Scrr$ and $\Scll$, do not
transform
under background gauge transformations.

With these definitions, performing the fermion integrations gives
(ignoring,
as always, a field-independent overall factor):
\label\fermfactor
\eq
\left[ {\det \Dps  \; \det \Dms \over \det \dps \; \det
\dms} \right]^{1/2} = \exp\left\{ -i \WZW{\Scll \, \ell \, r^\dagger \,
\Scrr^\dagger} \right\}.
\eeq

\topic{2. Changes of Variables I}

We next change variables from $A_\pm$ to $\Scll$ and $\Scrr$.  The
Jacobian of the transformation from $[dA_+]\,[dA_-]$ to the
group-invariant measure, $[d \Scrr]\,[d \Scll]$, is \pw:
\label\jacobian
\eq
J =  {\det \htDps  \; \det \htDms \over \det \dps \; \det \dms} =
 \exp\left\{ -i \kappa \WZW{\Scll \, \ell \, r^\dagger \, \Scrr^\dagger}
\right\},
\eeq
in which the `hat' over the gauge-covariant derivative is meant to
indicate that this derivative is to be taken in the adjoint representation:
$\widehat{D}_\pm X \equiv \partial_\pm X -i [ (\Scaa + A)_\pm , X]$. The
constant $\kappa$ here accounts for the difference in normalization
between the generators in the fundamental and the adjoint representations,
as well as for the absence of the overall square root of the determinants
in
eq.~\jacobian. If the adjoint generators, $T_a$, satisfy $\tr(T_a T_b) =
\lambda \, \delta_{ab}$, then $\kappa = 4 \lambda$.

\topic{3. Light-Cone Gauge}

We next choose to work within the background-covariant light-cone gauge,
$A_+ \equiv 0$. In terms of the Wilson-line variables we may implement
this gauge with the choice: $\Scrr  \equiv 1$, for which the
Lagrange-multiplier
term in eq.~\ggedgenfn\ simplifies considerably:
\label\lmterm
\eq \eqalign{
S_{\sss LM}  &\equiv 2 \int d^2x \, \tr \left( \Lambda V_{+-} \right) \cr
&= 2 \int d^2x \, \tr \left( \Lambda \htsdps A_- \right) \cr
&= -2i \int d^2x \, \tr \left[ (\htsdps \Lambda)
\, \ell^\dagger (\Scll^\dagger \dms \Scll) \, \ell \right], \cr}
\eeq
where the last equality requires an integration by parts. We have again
introduced a `hat' on the background-covariant derivative to emphasize
that it here acts in the adjoint representation.

\topic{4. Changes of Variables II}

\ref\newnad{E. Alvarez, L. Alvarez-Gaum\'e and Y. Lozano, preprint
CERN-TH-7204 (1994) (unpublished).}
The final step that is required in order to proceed is a judicious change
of
 variables for the Lagrange-multiplier field, $\Lambda$.\foot\thanks{
We thank Luis Alvarez-Gaum\'e for a key conversation on this point. A
general discussion of nonabelian duality using group-valued
dual variables may be found in Ref.~\newnad.}
In order to take advantage of the Polyakov-Wiegmann identity,
eq.~\pwidentity,
we wish to transform to a group-valued variable, $X$, for which we may
rewrite the quantity $\htsdps \Lambda$ in terms of the combination
$X^\dagger \dps X$. There are two considerations which can be used to
pin down the required change of variables: ($i$) Any relation between
$\htsdps \Lambda$ and $X^\dagger \dps X$ should be consistent with the
vectorlike background gauge invariance; and ($ii$) since $\htsdps
\Lambda$ is independent of the variable $\ell$, so must be $X$.

Given that background transformations take $\htsdps \Lambda$ into $\Scg
\, \htsdps \Lambda \, \Scg^\dagger$, we see that a background-covariant
choice for
the desired change of variables is:
\label\newvariable
\eq \eqalign{
\htsdps \Lambda &= i \xi \; r^\dagger (X^\dagger \dps X) \, r \cr
&= i \xi \; [ (Xr)^\dagger \dps (Xr)  - r^\dagger \dps r ] .\cr}
\eeq
Here $X$ is completely neutral under background gauge transformations,
with the usual transformation rule for $r$ -- \ie\ $r \to r \,
\Scg^\dagger$ --
providing the proper transformation
property for the right-hand side.  The parameter, $\xi$, is at present
an
arbitrary number which is to be chosen to simplify  later results.

With this choice, the Lagrange-multiplier term of eq.~\lmterm\ becomes:
\label\newlmterm
\eq \eqalign{
S_{\sss LM} &= 2\xi \int d^2x \, \tr \left\{ (X^\dagger \dps X) \, r
\ell^\dagger
 (\Scll^\dagger \dms  \Scll) \, \ell \, r^\dagger \right\} \cr
&= 2\xi \int d^2x \, \tr \left\{ (X^\dagger \dps X) \left[
 (\Scll \, \ell r^\dagger)^\dagger \dms  (\Scll \, \ell r^\dagger) -
  (\ell r^\dagger)^\dagger \dms  (\ell r^\dagger) \right] \right\} \cr
&= - 8 \pi \xi \, \bigl\{ \WZW{\Scll \, \ell r^\dagger X^\dagger} -
\WZW{\Scll \, \ell r^\dagger } - \WZW{\ell r^\dagger X^\dagger}
+ \WZW{\ell r^\dagger} \bigr\} . \cr}
\eeq
This final form follows after using the Polyakov-Wiegmann identity,
eq.~\pwidentity.

All that remains is to find the Jacobian for the
change of variables from
$\Lambda$ to $X$. For the purposes of doing so it is useful to think of
this
transformation as happening in two steps, first from $\Lambda$ to $\htsdps
\Lambda$, and then from $\htsdps \Lambda$ to $X$. The measures therefore
are
related by:
\label\naivej
\eq \eqalign{
[d \Lambda] &= [d(\htsdps \Lambda)] \left[ {\det \dps \over \det
\htsdps(r) }
\right] \cr
 &= [d X] \left[ {\det \htsdps(Xr) \over \det \htsdps(r) } \right],\cr}
\eeq
where the notation $\htsdps (g)$ is meant to indicate that the
corresponding
covariant derivative is constructed using the gauge field $ig^\dagger
\dps g$.
The Jacobian for transforming from $\htsdps \Lambda$ to $X$ may be
 recognized
as a special case of the Jacobian of eq.~\jacobian, for
transforming from a gauge potential to the corresponding
Wilson-line variable.

\ref\anomalies{See  for instance, L. Alvarez-Gaum\'e and P. Ginsparg,
{\it Ann. Phys.} (N.Y.) {\bf 161} (1985) 423.}
A problem presents itself as soon as one tries to make sense out of the
determinants which appear in eq.~\naivej. This is because these
determinants are
not yet unambiguously defined \anomalies, suffering as they do from an
anomaly for the
{\it vector} symmetry group, \von. As a result, although the original
measure,
$[d\Lambda]$, was supposed to be invariant under vectorlike background
gauge
transformations --- as is $[dX]$ trivially, since $X$ is invariant --- the
determinants which appear on the right-hand-side of eq.~\naivej\ are not.

We may remove these ambiguities by using the following two properties to
 more
completely define the determinants of interest. We firstly require that
both determinants be invariant under \von\ background gauge transformations.
Also, since lorentz-invariance would require an integration over
$\htsdms\Lambda$ as well as over $\htsdps\Lambda$,
we can think of eq.~\naivej\ as having been evaluated in a gauge for which
$\htsdms\Lambda= \dms \Lambda$. We therefore also require
that our \von-invariant result for the determinants agree with the naive
application of eq.~\answer\ to eq.~\naivej\ in this gauge.

These two conditions introduce an $\ell$-dependence into the result, and
uniquely specify the determinants to be:
\label\realj
\eq [d \Lambda] =  [d X] \exp \left\{ i \kappa \, \WZW{\ell \, r^\dagger}
- i \kappa \WZW{\ell \, r^\dagger X^\dagger} \right\}.
\eeq
\endtopic

We may now put the above four results together to simplify our starting
expression, eq.~\ggedgenfn. We have:
\label\simplification
\eq \eqalign{
Z[\Scaa] &= \int [d\psi] \, [dA_+] \, [dA_-] \, [d\Lambda] \;
\exp \left\{ i \int
d^2x \;  \left[ - \, \hf \; \psibar \,  \left( \gamma^+ \Pr \Dps +
 \gamma^- \Pl
\Dms  \right) \psi \right. \right.  \cr
& \left. \left.
\phantom{\int [d\psi] \, [dA_+] \, [dA_-] \, [d\Lambda] \;
\exp  i \int d^2x \;
 - \, \hf \; \psibar }
+ 2 \,  \tr\left( \Lambda
V_{+-} \right) \right] \right\} \; \delta[A_+] \cr
&= \int [dX] \, [d\Scll] \; \exp \Bigl\{ -i (1 + \kappa) \WZW{\Scll \,
\ell
r^\dagger} + i \kappa \Bigl( \WZW{\ell \, r^\dagger} - \WZW{\ell
\, r^\dagger
X^\dagger} \Bigr)  \cr
& \phantom{\int [dX] \, [d\Scll] \; \exp \Bigl\{ }
 - 8 \pi i \xi \, \Bigl(  \WZW{\Scll \, \ell r^\dagger X^\dagger} -
\WZW{\Scll \, \ell r^\dagger } - \WZW{\ell r^\dagger X^\dagger}
+ \WZW{\ell r^\dagger} \Bigr)  \Bigr\}. \cr}
\eeq
Each of the terms in this last expression correspond
to one of the items from
the above discussion\foot\deltafunction{
Notice that implicit in equation~\simplification\ there is a nontrivial
path integral representation of the $\delta$--function as a result
of the change of variables II.}. To wit: (1) the factor proportional to
$(1+\kappa)$ originates from the fermion determinant, and the
Jacobian for
the change of variables from $A_\pm$ to $\Scrr$ and $\Scll$
(eqs.~\fermfactor\ and
\jacobian); (2) the terms proportional to $\kappa$ arise
due to the Jacobian of
eq.~\realj\ for the change from $\Lambda$ to $X$; and (3) the terms
proportional
to $\xi$ correspond to the Lagrange-multiplier lagrangian of eq.~\newlmterm.

At this point it is worthwhile to make a helpful choice for the parameter
$\xi$. Since the coefficient which premultiplies the factor $\WZW{\Scll \,
\ell
r^\dagger}$ is proportional to $(1 + \kappa - 8 \pi \xi)$, it is
irresistible
to choose $8 \pi \xi = 1+\kappa$, in which case this entire term
vanishes. This has the great advantage of completely decoupling
the $[d\Scll]$
integration from all of the others, since $\Scll$ then only enters the
functional
integrand through the overall multiplicative factor
\label\factor
\eq
\int [d\Scll] \exp \Bigl\{ -i (1+\kappa) \WZW{\Scll \, \ell r^\dagger
X^\dagger}
\Bigr\}.
\eeq
The change of variables $\Scll \to \widehat{\Scll} = \Scll \,
\ell r^\dagger
X^\dagger$, for which $[d\widehat{\Scll}] = [d\Scll]$, then shows
this to be an
irrelevant field-independent constant.

We are left with
\label\almostdone
\eq  \eqalign{
Z[\Scaa] &= \int [dX] \; \exp \Bigl\{ i (\kappa - 8 \pi \xi)
\Bigl( \WZW{\ell
\, r^\dagger} - \WZW{\ell \, r^\dagger X^\dagger} \Bigr) \Bigr\} \cr
&= \int [dX] \; \exp \Bigl\{ -i \Bigl( \WZW{\ell \, r^\dagger} -
\WZW{\ell \,
r^\dagger X^\dagger} \Bigr) \Bigr\} . \cr}
\eeq
This may be put into a more familiar form simply by redefining fields
$X \to g$,
with $X^\dagger = r \, g \, r^\dagger$, for which $[dX] = [dg]$.
With this choice $g$ inherits the transformation rule $g \to \Scg \, g \,
\Scg^\dagger$ under the vectorlike symmetry. The final, bosonized, result
becomes:
\label\bosonized
\eq
 Z[\Scaa] = \int [dg] \; \exp \Bigl\{ i \Gamma_{\sss GWZW}(g,\Scaa) \Bigr\},
\eeq
where
\label\gwzwformula
\eq \eqalign{
\Gamma_{\sss GWZW}(g,\Scaa) &= \WZW{\ell \, g \, r^\dagger} - \WZW{\ell \,
r^\dagger} \cr
&= \WZW{g} + {1 \over 4 \pi} \int d^2x \, \tr \left[
i g^\dagger \dms g \, \Scaa_+ - i \dps g \, g^\dagger  \, \Scaa_- + g \,
\Scaa_+
g^\dagger \Scaa_- - \Scaa_+ \Scaa_- \right] . \cr}
\eeq
\ref\gwzwaction{D. Karabali and H.J. Schnitzer, \npb{329}{90}{649},
 and references cited therein.}
This second way of writing the bosonized action re-expresses the
Wilson-line
variables $\ell$ and $r$ in terms of the original fields, $\Scaa_\mu$.
In this form
it may be recognized as the gauged Wess-Zumino-Witten action,
which is usually derived by `gauging' the global \von\
symmetry of $\WZW{g}$, using \eg\ the Noether prescription \gwzwaction.
This action also properly reproduces the fermion anomaly under the chiral
\chon\ group provided that $g$ is defined to transform in the standard way:
$g \to \Scl \, g \, \Scr^\dagger$.

\ref\pimethods{P. di Vecchia, B. Durhuus and J.L. Petersen,
\plb{144}{84}{245};\bk
D. Gonzalez and A.N. Redlich, \plb{147}{84}{150}.}
The expression for $\Gamma_{\sss GWZW}$ in terms of Wilson-line variables
is useful in that it makes clear that the remaining integration over $g$ can
be explicitly performed. Changing variables to $\hat{g} = \ell \, g \,
r^\dagger$
(with $[d\hat{g}] = [dg]$) shows that the integration over $\hat{g}$ simply
provides an overall field-independent normalization constant.
We arrive in this way back to our expected result,
eq.~\answer, for $Z[\Scaa]$. Notice also that by looking at the correlation
functions, differentiating $Z[\Scaa]$ with respect to the background
fields $a_{\pm}$ evaluated at $a_{\pm}=0$ we find the known
correspondence among the fermionic and the bosonic currents:
\label\currents\eq
i\psibar \gamma_- \psi \leftrightarrow {i\over 2{\pi}}\, g^\dagger
\dms g , \qquad \hbox{and:} \qquad
i\psibar \gamma_+ \psi \leftrightarrow
- \, {i\over 2\pi}\, \dps g \, g^\dagger  ,\eeq
together with the delta-function contact terms in the correlations between
the left- and right-handed currents that had been found
elsewhere using path-integral methods \pimethods.

Finally, following the same arguments as in ref.~\basd, it is
also possible
to extend this analysis to include four fermion couplings as well as
mass terms, reproducing the results of ref.~\witten.

\vfill\eject

\section{Conclusions}

Our purpose has been to provide a constructive and systematic derivation
of the rules for nonabelian bosonization that were first written down by
Witten
some ten years ago. We have done so by showing that nonabelian bosonization
 may
be considered to be a special case of a nonabelian duality transformation,
for
which systematic and constructive techniques have recently been formulated.
It
is our hope that this connection can lead to a wider application and
understanding of both nonabelian bosonization and duality transformations.

\centerline{\bf Acknowledgments}

\bigskip

We would like to acknowledge helpful conversations with Luis
Alvarez-Gaum\'e, Abhijit Kshirsagar and Tim Morris. We also
thank Rob Myers for a critical reading of a preliminary version of
this paper. This research
was partially funded by N.S.E.R.C.\ of Canada,
les Fonds F.C.A.R.\ du Qu\'ebec, and by the Swiss National Foundation.

\listrefs

\bye